\documentclass[aps,pra,twocolumn,superscriptaddress,nofootinbib]{revtex4-2}
\usepackage{graphics}
\usepackage{amsmath}
\usepackage{graphicx}
\usepackage{color}
\usepackage{amsfonts}
\usepackage{amssymb}
\usepackage{float}
\usepackage{tikz}
\usetikzlibrary{positioning}
\usetikzlibrary{backgrounds}
\usepackage{caption}
\usepackage{subcaption}
\usepackage{bm}
\usepackage{hyperref}

\newtheorem{theorem}{Theorem}
\newtheorem{corollary}{Corollary}[theorem]
\newtheorem{lemma}[theorem]{Lemma}

\begin{document}

\title{A no-go theorem for privacy in distributed sensing using Gaussian states}

\author{Jason L. Pereira} \email{Jason.Pereira@lip6.fr}
\affiliation{Sorbonne Universit\'{e}, CNRS, LIP6, 4 Place Jussieu, Paris F-75005, France}
\author{Damian Markham}
\affiliation{Sorbonne Universit\'{e}, CNRS, LIP6, 4 Place Jussieu, Paris F-75005, France}

\date{\today}

\begin{abstract}
    In the discrete variable setting, entangled resource states allow a set of parties to learn a global function of a set of spatially separated systems, whilst keeping the local parameters of those systems completely private. In the continuous variable setting, distributed sensing has been carried out using Gaussian resource states, but without the same guarantees about privacy. Here, we show that perfect privacy is impossible to achieve for any distributed sensing protocol that uses Gaussian states as a resource. We also introduce a measure of relative privacy, bounding the degree to which any Gaussian distributed sensing protocol can keep local parameters hidden.
\end{abstract}

\maketitle

\section{Introduction}

Distributed sensing is the task of estimating a global function of parameters of remote systems. A set of parties carry out quantum metrology on a set of spatially separated systems, each characterised by some individual parameter, in order to learn a function of the parameter values. Using shared quantum resources, it is possible to outperform any separable scheme, in which the parties do not have access to entangled probe systems. Specifically, for certain sensing problems, parties with access to a shared graph state can achieve enhanced sensitivity~\cite{proctor_multiparameter_2018,zhuang_distributed_2018,ge_distributed_2018,shettell_graph_2020} and privacy~\cite{shettell_private_2022,hassani_privacy_2025,bugalho_private_2025}.

Enhanced sensitivity means that the precision with which the parties can learn the target function scales better with the number of parties than could be achieved by any strategy that estimates the individual, local parameter values and then calculates the value of the function using the estimates. This is true even if the local strategy uses entanglement with an idler system.

Privacy means it is not possible to gain any more information about the local parameters than is contained in the value of the target function. This should hold even if some parties are dishonest. We can also phrase this as ``any party can hide the value of any other party's local parameter".

Whether these properties can be achieved depends on the shared resource state that the parties have access to. In the discrete variable (DV) setting, it is known that any resource that is equivalent to a GHZ state up to local unitaries enables both enhanced sensitivity and perfect privacy for certain sensing problems~\cite{eldredge_optimal_2018,bugalho_private_2025,hassani_privacy_2025}.

In the continuous variable (CV) setting, distributed sensing protocols have been proposed and implemented using Gaussian graph states~\cite{gatto_distributed_2019,guo_distributed_2020}. These demonstrate enhanced sensitivity (defined in terms of local strategies with probes of the same energy), but do not have privacy in the same sense as in the DV case~\cite{junior_privacy_2025}. Therefore, a natural question is whether there can exist a distributed sensing protocol, using Gaussian states as a resource, that can demonstrate both privacy and enhanced sensitivity. Ref.~\cite{junior_privacy_2025} addressed this question for specific classes of Gaussian state, but it has not yet been answered in general.

In this work, we will definitively answer the question of whether there exists a distributed sensing protocol, using Gaussian resource states, that has perfect privacy in the negative. We will present a necessary condition for the existence of a private encoding and will show that this condition cannot be met by Gaussian states. We will then briefly discuss how privacy can be simply restored through sharing a classical secret, before finally addressing how we can bound a measure of relative (i.e., not perfect) privacy for Gaussian resource states.

\section{Setting}

\subsection{Private, distributed sensing}

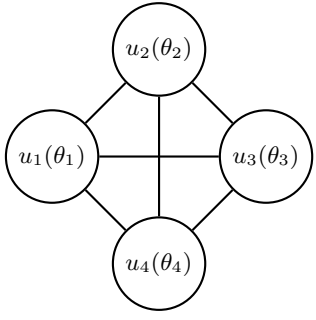
\begin{figure}[t]
    \centering
    \begin{tikzpicture}[node distance={20mm}, thick, main/.style = {draw, circle}]
        \node[main] (1) {$u_1(\theta_1)$};
        \node[main] (2) [above right of=1] {$u_2(\theta_2)$};
        \node[main] (3) [below right of=2] {$u_3(\theta_3)$};
        \node[main] (4) [below right of=1] {$u_4(\theta_4)$};

        \draw (1) -- (2);
        \draw (1) -- (3);
        \draw (1) -- (4);
        \draw (2) -- (3);
        \draw (2) -- (4);
        \draw (3) -- (4);
        
    \end{tikzpicture}
    \caption{The basic setup for a distributed sensing problem. For privacy, we want to choose the parametrised unitaries, $u_i$, so that the encoded state only holds information about a linear function of the parameters, and not their individual values.}
    \label{fig: scenario}
\end{figure}

We consider $N$ distinct parties who share a resource state, $| \psi \rangle$, that they want to use for private, distributed sensing. We are primarily interested in graph states, but will allow $| \psi \rangle$ to be any $N$-partite state. For simplicity, we start by considering unitary encoding channels and pure resource states. We assume individual, separable encodings, so that each party has a parameter, $\theta$, and the overall encoding unitary can be written as $U_{\mathrm{enc}}(\{\theta_i\}) = \bigotimes_{i=1}^N u_i(\theta_i)$.
Both the encoding unitary and the encoded parameter are party specific, so that different parties could have different encoding unitaries (e.g., in the DV case, party $1$ could be encoded with $e^{i \theta_1 X}$, whilst party $2$ could be encoded with $e^{i \theta_2 Z}$). This situation is illustrated in Fig.~\ref{fig: scenario}

Refs.~\cite{shettell_private_2022,bugalho_private_2025} define a \textit{private encoding} as one for which the encoded state, $|\psi_{\mathrm{enc}}(\{\theta_i\})\rangle = U_{\mathrm{enc}}(\{\theta_i\}) | \psi \rangle$, only holds information about a single function (linear combination), $f(\{\theta_i\})$, of the set of parameters. Ref.~\cite{bugalho_private_2025} also defines a private state as one that admits a private encoding for some set of encoding unitaries, and defines a privacy measure based on the quantum Fisher information (QFI) matrix of the encoded state. Perfect privacy is when this matrix has only one non-zero eigenvalue.
We can also extend our definition to encoding channels, rather than just encoding unitaries.

By definition, if an encoding is private, there exists no protocol by which information about any other linear combination of parameters can be retrieved from the encoded state. Equivalently, given access to the encoded state, no subset of dishonest parties can retrieve any information about the individual parameter values of the remaining parties that is not already contained in the value of the target function, as evaluated by the honest parties.
We can also phrase this as: any two honest parties can hide each other's parameter values, in the sense that it is impossible to know which of the parties encoded the state, in order to achieve a particular value of the target function.

\subsection{Resource states and graph states}

Any pre-shared (between the $N$ parties) quantum state is called a resource state, since it is a resource they can use for a task. We will often speak generically about resource states, but we are sometimes specifically interested in a particular class of resources called graph states~\cite{hein_entanglement_2006}.

In DV settings, these can be defined as states that are constructed in a specific way, so that they can be represented by an undirected graph. Namely (for qubits): each node of the graph represents a qubit prepared in the $|+\rangle$ state, and edges are implemented by applying a controlled-$Z$ operation between the two nodes connected by that edge. They can be equivalently defined in terms of stabilisers.

We are particularly interested in the CV setting and, specifically, Gaussian resources. In this setting, graph states are less consistently defined, with various definitions being used. For instance, the stabiliser formalism only applies in the limit of infinite energy~\cite{gu_quantum_2009,menicucci_graphical_2011}. For our purposes, we again use a definition based on how they are constructed. Specifically, each node represents a vacuum mode to which we have applied a one-mode squeezing operation ($\exp[i \nu (\hat{a}^2 - \hat{a}^{\dagger 2})]$) and edges are created by applying a controlled-phase gate ($\exp[i \eta \hat{q}_1 \hat{q}_2]$) between two nodes. Notably, both types of operation involve a strength parameter ($\nu$ and $\mu$) so the graphs are no longer unweighted; we do not require that $\nu$ is the same for every node or that $\mu$ is the same for every edge. We also allow local (to each node) Gaussian operations after all controlled-phase gates have been applied, since they do not affect the entanglement properties.

\section{Necessary conditions for private encodings to exist}\label{sec: conditions}

To have any hidden information, we require that there exist at least two sets of parameters, $\{\theta^{(a)}_i\}$ and $\{\theta^{(b)}_i\}$, that result in the same encoded state. Otherwise, every encoded state, $|\psi_{\mathrm{enc}}(\{\theta_i\})\rangle$, would be one-to-one with exactly one set of individual parameter values, so that, with the correct observable, one could learn all of the $\theta_i$. Hence, we must have $U_{\mathrm{enc}}(\{\theta^{(a)}_i\}) | \psi \rangle = U_{\mathrm{enc}}(\{\theta^{(b)}_i\}) | \psi \rangle$, and so $U^{\dagger}_{\mathrm{enc}}(\{\theta^{(b)}_i\}) U_{\mathrm{enc}}(\{\theta^{(a)}_i\}) | \psi \rangle = | \psi \rangle$. Since, $U_{\mathrm{enc}}$ is separable, there must be a separable unitary that acts on $| \psi \rangle$ as the identity, in order for there to be any hidden information about the parameters $\{\theta_i\}$.

For privacy, we have the stronger condition that we cannot learn any function of the parameters other than $f$. This means we are not allowed to learn the difference in parameter values between any pair of (honest) parties. This is only possible if, for every pair of parties $(i,j)$, there exist (non-trivial, i.e., not both equal to $0$) parameters $(\theta_i,\theta_j)$ such that $|\psi_{\mathrm{enc}}(\{0,...,0,\theta_i,0,...,0,\theta_j,0,...0\})\rangle = | \psi \rangle$, where, without loss of generality, we have set every other parameter to $0$. I.e., there must exist a pair of unitaries, $u$ and $v$, such that $u_i \otimes v_j \otimes \mathbf{I}_{\overline{ij}}$ acts as the identity on $| \psi \rangle$, where $\mathbf{I}_{\overline{ij}}$ denotes the identity on all modes except for $i$ and $j$ and where $u$ and $v$ are not both equal to the identity.

\begin{lemma}
    Resource state $\Psi$ can only admit a private, unitary encoding if, for every $i$ and $j$, there exist unitaries $u$ and $v$ such that:
    \begin{enumerate}
        \item $(u_i \otimes v_j \otimes \mathbf{I}_{\overline{ij}}) \Psi (u_i^{\dagger} \otimes v_j^{\dagger} \otimes \mathbf{I}_{\overline{ij}}) = \Psi$,
        \item $(u_i \otimes \mathbf{I}_{\overline{i}}) \Psi (u_i^{\dagger} \otimes \mathbf{I}_{\overline{i}}) \neq \Psi$.
    \end{enumerate}
    \label{lemma: all party condition}
\end{lemma}

We use $\Psi$ instead of $| \psi \rangle$ to make clear that the lemma holds for density matrices as well as pure states.
Condition 2 ensures the encoding unitaries actually encode information, and stops us from defining $u$ as a global phase or as acting only on an unsupported subspace.

The following (trivial) observation allows us to derive a single party, necessary condition from Lemma~\ref{lemma: all party condition}:
\begin{equation}
    \begin{split}
        \Psi &= (u_i \otimes v_j \otimes \mathbf{I}_{\overline{ij}}) \Psi (u^{\dagger}_i \otimes v^{\dagger}_j \otimes \mathbf{I}_{\overline{ij}})\\
        &\implies \mathrm{Tr}_{\overline{i}}[\Psi] = u_{i} \mathrm{Tr}_{\overline{i}}[\Psi] u^{\dagger}_{i}.
    \end{split}
    \label{eq: 1 party trace condition}
\end{equation}
This expresses the fact that the encoding unitary must act as the identity on each local state. Otherwise, one could learn each parameter, $\theta_i$, by ignoring all the other parties and carrying out a single party measurement.

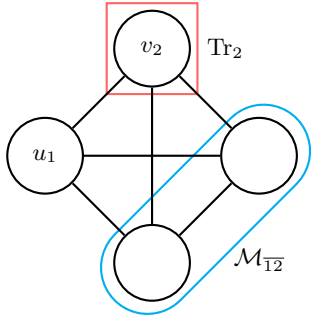
\begin{figure}[t]
    \centering
    \begin{tikzpicture}[node distance={20mm}, thick, main/.style = {draw, circle, minimum size=10mm}, measurement/.style={line join=round, double,line cap=round,double distance=#1,shorten >=-#1/2,shorten <=-#1/2, thick}, measurement/.default=12mm, trace/.style = {rectangle, draw=red!60, thick, minimum size=12mm}]
        \node[main] (1) {$u_1$};
        \node[main] (2) [above right of=1] {$v_2$};
        \node[main] (3) [below right of=2] {};
        \node[main] (4) [below right of=1] {};

        \draw (1) -- (2);
        \draw (1) -- (3);
        \draw (1) -- (4);
        \draw (2) -- (3);
        \draw (2) -- (4);
        \draw (3) -- (4);

        \begin{scope}[on background layer]     
            \draw[measurement, cyan] (3) -- (4);
            \node[trace] () [above right of=1] {};
            \node[below=6mm of 3] () {$\mathcal{M}_{\overline{12}}$};
            \node[right=1mm of 2] () {$\mathrm{Tr}_{2}$};
        \end{scope}
        
    \end{tikzpicture}
    \caption{An illustration of the necessary condition for privacy stated in Theorem~\ref{theorem: general condition}. If there exists a pair of unitaries, $u$ and $v$, that collectively act as the identity, when applied to nodes $1$ and $2$, then $u$ must act as the identity on all possible conditional states after tracing out node $2$ and carrying out a measurement on all nodes except for $1$ and $2$.}
    \label{fig: theorem}
\end{figure}

In fact, Lemma~\ref{lemma: all party condition} lets us derive a set of other single party, necessary (not sufficient) conditions. We start by introducing some more notation. Let $\mathcal{M}_{\overline{ij}}$ denote some measurement on all modes except for $i$ and $j$, and let $\mathcal{M}^{(m)}_{\overline{ij}}\big\{\rho\big\}$ denote the conditional state of the remaining modes of $\rho$ after carrying out measurement $\mathcal{M}_{\overline{ij}}$ and receiving measurement outcome $m$. Then,
\begin{equation}
    \begin{split}
        \Psi &= (u_i \otimes v_j \otimes \mathbf{I}_{\overline{ij}}) \Psi (u^{\dagger}_i \otimes v^{\dagger}_j \otimes \mathbf{I}_{\overline{ij}})\\
        &\implies \mathrm{Tr}_{j}[\Psi] = (u_i \otimes \mathbf{I}_{\overline{ij}}) \mathrm{Tr}_{j}[\Psi] (u^{\dagger}_i \otimes \mathbf{I}_{\overline{ij}})\\
        &\implies \mathcal{M}^{(m)}_{\overline{ij}}\big\{\mathrm{Tr}_{j}[\Psi]\big\} = u_i \mathcal{M}^{(m)}_{\overline{ij}}\big\{\mathrm{Tr}_{j}[\Psi]\big\} u^{\dagger}_i,
    \end{split}
    \label{eq: 1 party conditional condition}
\end{equation}
where $u_i$ commutes with the measurement, since they act exclusively on different modes. We therefore need a single unitary, $u$, that acts as the identity on \textit{every} conditional state $\mathcal{M}^{(m)}_{\overline{ij}}\big\{\mathrm{Tr}_{j}[\Psi]\big\}$ (for every measurement, $\mathcal{M}_{\overline{ij}}$, and outcome, $m$).
This condition is illustrated in Fig.~\ref{fig: theorem}.

For $u$ to act as the identity on some state, $\rho$, $u$ must be diagonalisable in the same basis as $\rho$. If and only if this is the case, it can apply a relative phase to different eigenvectors of $\rho$, whilst leaving $\rho$ unchanged. Since we require a single $u$ to act as the identity on every state $\mathcal{M}^{(m)}_{\overline{ij}}\big\{\mathrm{Tr}_{j}[\Psi]\big\}$, every such state must have at least one shared eigenvector.

\begin{theorem}
    Let $\Psi$ be a multimode state shared between $N>2$ parties. Then, a private unitary encoding cannot exist unless, for every pair of parties $(i,j)$, there exists an eigenvector of $\mathcal{M}^{(m)}_{\overline{ij}}\big\{\mathrm{Tr}_j[\Psi]\big\}$ that is the same for every measurement $\mathcal{M}_{\overline{ij}}$ and every measurement outcome $m$.
    \label{theorem: general condition}
\end{theorem}

This condition is met by DV GHZ-states (and states that can be reached from a GHZ-state via local unitaries), since tracing over any single party results in a fully unentangled state. On the other hand, given any other DV graph state, there exists a choice of $(i,j)$ such that the condition in Theorem~\ref{theorem: general condition} cannot be met. We will also show it can never be met by a Gaussian state.

As a simple corollary, we might also consider a weaker definition of privacy, where we require every member of any clique of $n<N$ parties to be able to hide each other's values. For example, we might believe that for some $| \psi \rangle$, not every pair of parties $(i,j)$ can hide one another's values, but that every triple of parties $(i,j,k)$ can collaborate to hide one another's parameter values. Physically, this would refer to a situation in which we are guaranteed to have at least $n>2$ trusted parties, but do not know their identities (in contrast to the previous scenario, in which we set $n=2$). Theorem~\ref{theorem: general condition} can be extended to this more general situation.

\begin{corollary}
    Let $\Psi$ be a multimode state shared between $N$ parties. A subset of parties, $J$, cannot hide one another's unitary-encoded parameter values unless, for each $i\in J$, there exists an eigenvector of $\mathcal{M}^{(m)}_{\overline{J}}\big\{\mathrm{Tr}_{J\setminus i}[\Psi]\big\}$ that is the same for every measurement $\mathcal{M}_{\overline{J}}$ and every measurement outcome $m$.
    \label{theorem: clique condition}
\end{corollary}

Since we will show that the condition in Corollary~\ref{theorem: clique condition} also cannot be met for Gaussian states, we may consider an even weaker definition of privacy. For an encoding to be private, the encoded state must carry no information about the hidden parameter values. However, in practice, the dishonest parties do not have access to the entire encoded state, just their own modes and the measurement results of the honest parties. We might, in theory, consider a \textit{private protocol}, wherein the encoded state holds information about the individual parameter values of the honest parties, but the measurement statistics of the honest parties and the conditional states of the dishonest parties do not.
We can see this as a particular case of finding a private encoding for a set of non-unitary encoding channels, since we can treat the measurement as a quantum to classical channel.

Hence, as a second corollary, let us now consider non-unitary encoding channels. We now allow the encoding to be enacted by parametrised channels, $\mathcal{E}_i(\theta_i)$, rather than by parametrised unitaries. In the honest case, the encoded state is then $\Psi_{\mathrm{enc}} = \big( \bigotimes_{i=1}^N \mathcal{E}_i(\theta_i) \big)[\Psi]$. If a subset of parties are being dishonest, they might choose to act with the identity, rather than a channel. For privacy, we require that for every pair of parties $(i,j)$ and for every choice of parameter $\theta_i$, there exists some choice of $\theta_j$ such that $\big( \mathcal{E}_i(\theta_i) \otimes \mathcal{E}_j(\theta_j) \otimes \mathbf{I}_{\overline{ij}} \big)[\Psi] = \big( \mathcal{E}_i(0) \otimes \mathcal{E}_j(0) \otimes \mathbf{I}_{\overline{ij}} \big)[\Psi]$. This is a slight generalisation of Lemma~\ref{lemma: all party condition}, allowing for the fact that the encoding channel does not necessarily act as the identity, even for the case in which $\theta_i = 0$. Following the same line of reasoning as in the unitary encoding case, we can arrive at a necessary condition for a channel encoding to be private.

\begin{corollary}
    Let $\Psi$ be a multimode state shared between $N$ parties. A subset of parties, $J$, cannot hide one another's channel-encoded parameter values unless, for each $i\in J$, there exists a parametrised set of channels, $\{\mathcal{E}_i(\theta)\}$, such that $\mathcal{E}_i(\theta)\Big[\mathcal{M}^{(m)}_{\overline{J}}\big\{\mathrm{Tr}_{J\setminus i}[\Psi]\big\}\Big]$ is the same for every value of $\theta$, every measurement $\mathcal{M}_{\overline{J}}$, and every measurement outcome $m$, but $\mathcal{E}_i(\theta)\otimes\mathbf{I}_{\overline{i}}[\Psi]$ is not the same for every value of $\theta$.
    \label{theorem: channel condition}
\end{corollary}

By showing that the conditions in Theorem~\ref{theorem: general condition} and Corollaries~\ref{theorem: clique condition} and \ref{theorem: channel condition} cannot be met by any Gaussian resource, we can show that a distributed sensing protocol using Gaussian states can never achieve perfect privacy.

\section{No-go on perfect privacy for Gaussian states}

For Gaussian states, carrying out a Gaussian measurement on part of a state results in a Gaussian state with a covariance matrix that does not depend on the measurement result and a first moments vector that does. Specifically, let the initial covariance matrix, $V_{\mathrm{init}}$, and first moments vector, $X_{\mathrm{init}}$, take the forms
\begin{equation}
    V_{\mathrm{init}} = \begin{pmatrix}
        \mathbf{A} &\mathbf{C}\\
        \mathbf{C}^T &\mathbf{B}
    \end{pmatrix},
    \quad X_{\mathrm{init}} = \begin{pmatrix}
        X_A\\
        X_B
    \end{pmatrix},
    \label{eq: cov and fmv}
\end{equation}
where $\mathbf{A}$ and $X_A$ are the covariance matrix and first moments vector of the system to be measured, $\mathbf{B}$ and $X_B$ are the covariance matrix and first moments vector of the unmeasured system, and $\mathbf{C}$ are the cross-correlations in the covariance matrix. For a measurement outcome of $\mathbf{o}$, the covariance matrix, $V^{(\mathbf{o})}$, and first moments vector, $X^{(\mathbf{o})}$, of the resulting conditional state take the forms
\begin{align}
    &V^{(\mathbf{o})} = \mathbf{B}-\mathbf{C}^T (\mathbf{A}+\mathbf{M})^{-1} \mathbf{C},\\
    &X^{(\mathbf{o})} = X_B + \mathbf{C}^T (\mathbf{A}+\mathbf{M})^{-1} (\mathbf{o}-X_A),\label{eq: X displacement}
\end{align}
where $\mathbf{M}$ is the covariance matrix of the measurement, i.e., the covariance matrix of the state onto which we are projecting the measured system. In the case of a homodyne measurement, this is the covariance matrix of an infinitely squeezed state and for a heterodyne measurement, it is the covariance matrix of a coherent state.
As a particular example, for Gaussian graph states, measuring a node in the $\hat{q}$ basis removes it from the graph and imposes some displacements on the neighbouring nodes.

Now let us consider the form of the state $\mathrm{Tr}_j[\Psi]$ or $\mathrm{Tr}_{J\setminus i}[\Psi]$, from Theorem~\ref{theorem: general condition} and Corollaries~\ref{theorem: clique condition} and \ref{theorem: channel condition}.
For convenience, we will use the latter form, since it is more general.
For any fixed measurement on the $\overline{J}$ modes, we can find an outcome-independent, diagonalising, Gaussian unitary such that the conditional output states, $\mathcal{M}^{(m)}_{\overline{J}}\big\{\mathrm{Tr}_{J\setminus i}[\Psi]\big\}$, are displaced thermal states.

If we pick that the measurement we carry out, $\mathcal{M}_{\overline{J}}$, is a sequence of heterodyne measurements on each $\overline{J}$ mode, every possible displacement to the first moments of the unmeasured mode can be enacted by a particular measurement outcome. Specifically, consider the expression for the displacement of the conditional state in Eq.~(\ref{eq: X displacement}). Identifying the $A$ system with the measured modes and the $B$ system with mode $i$, the displacement of mode $i$ is given by $\mathbf{C}^T (\mathbf{A}+\mathbf{I})^{-1}$ multiplied by the measurement outcome (where $\mathbf{M}=\mathbf{I}$, since we are heterodyning). As long as the two rows of $\mathbf{C}^T (\mathbf{A}+\mathbf{I})^{-1}$ are not proportional and neither have all of their elements set to $0$, we are always able to find a vector $\mathbf{o}$ such that $\mathbf{C}^T (\mathbf{A}+\mathbf{I})^{-1} (\mathbf{o}-X_A)$ can take any value we choose. If the two rows \textit{are} proportional to each other, we do not have two linearly independent displacements, so will always be displacing our conditional state along the same line, for any outcome, and if one of the rows has all of its elements set to $0$, we are only be able to induce displacements in one direction ($\hat{q}$ or $\hat{p}$). However, if this is the case, we can always rectify this by applying a Gaussian unitary to the $\overline{J}$ modes prior to measurement.\footnote{To verify this, recall that we can always find a Gaussian unitary that diagonalises $\mathbf{A}$, so we can assume $\mathbf{A}$ is diagonal without loss of generality. Then, a phase rotation on a single mode will not change $(\mathbf{A}+\mathbf{I})^{-1}$ but will rearrange the values of a $2$ by $2$ block of $\mathbf{C}^T (\mathbf{A}+\mathbf{I})^{-1}$, so we can always make it so that the two rows of $\mathbf{C}^T (\mathbf{A}+\mathbf{I})^{-1}$ are not proportional to each other and that neither has all of its elements set to $0$, as long as the elements of $\mathbf{C}$ are not all $0$. If all elements of $\mathbf{C}$ are $0$, the $\overline{J}$ modes are uncorrelated with the $i$ mode, and so cannot induce a displacement. This can only be the case for every clique $J$ if $\Psi$ is completely uncorrelated between modes, so no distributed sensing is possible anyway.}

To meet the condition in Theorem~\ref{theorem: general condition} or Corollary~\ref{theorem: clique condition}, we would therefore need a single state to be an eigenvector of both the undisplaced thermal state and that thermal state with every possible displacement applied. This cannot be the case: for zero displacement, the eigenvectors of a thermal state are the Fock states, and displacements never map one Fock state onto another.

Similarly, to meet the condition in Corollary~\ref{theorem: channel condition}, we would need there to exist (at least) two different channels, $\mathcal{E}_1$ and $\mathcal{E}_2$, that act identically on (give the same output for) every displaced thermal state input that mode $i$ can be projected onto. However, we note that the channels must act identically on mode $i$ for every choice of measurement, so we can also choose $\mathbf{M}=k\mathbf{I}$ for some $k>1$, which would lead to displaced thermal states with different (a continuous range of) average photon numbers. This condition can therefore only be fulfilled if the channels act identically on every displaced number state. Since the displaced vacuum states (i.e., coherent states) form an overcomplete basis, so that every input can be written as a quasiprobability distribution over them, if two channels act the same on all such states, they must therefore give the same output for every input. However, this means that they are, by definition, the same channel.

The conditions in Theorem~\ref{theorem: general condition} and Corollaries~\ref{theorem: clique condition} and \ref{theorem: channel condition} therefore cannot be met by any Gaussian state, and hence Gaussian states cannot admit private protocols.
As a clarification, this does not mean that it is impossible for \textit{any} pair of parties to hide each other's parameter values, but rather that it cannot hold that \textit{every} party is able to hide every other party's values.

\section{Restoring privacy with shared secret information}

If the parties have pre-shared secret information, they can restore privacy in what might be considered a trivial way. Specifically, if each party has access to a value, $\xi_i$, known only to them (not to the other parties), such that $\sum_{i=1}^N \xi_i = 0$, they can encode the value $(\theta_i+\xi_i)$ instead of just $\theta_i$ (e.g., by applying $u_i(\xi_i)$ before applying the encoding unitary in the normal way with their unknown secret parameter, $u_i(\theta_i)$). A protocol that estimates the sum of the parameters will then find the value $\sum_{i=1}^{N}(\theta_i+\xi_i) = \sum_{i=1}^{N}\theta_i$, so will still give the correct value. However, any scheme that lets dishonest parties learn the parameter value of party $i$ will estimate $(\theta_i+\xi_i)$ instead of $\theta_i$. Without access to every other part of the secret, i.e., every $\xi_j$ for $j\in\overline{i}$, it is not possible to recover $\xi_i$, and so it is not possible to learn anything about $\theta_i$ (so long as $\xi_i$ is uniformly distributed over the same range as $\theta_i$). Thus, just as in the case of private, distributed sensing with GHZ-states, as long as at least two parties are honest, no set of dishonest parties can learn anything about their individual parameter values. The shared secret could be established from pairwise secrets; specifically, if every pair of parties $(i,j)$ shares a pair of values $(\chi_{ij},\chi_{ji})$ such that $\chi_{ij}=-\chi_{ji}$, party $i$ can construct the value $\xi_i$ as $\xi_i=\sum_{j=1}^N \chi_{ij}$.

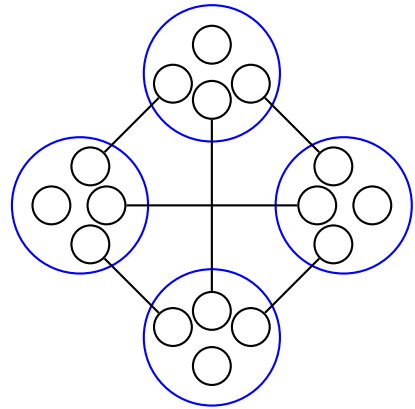
\begin{figure}[t]
    \centering
    \begin{tikzpicture}[node distance={30mm}, thick, main/.style = {draw, circle, minimum size=5mm}, party/.style = {draw=blue, thick, circle, minimum size=18mm}]
        \node[main] (10) {};
        \node[main] (12) [above right=2mm of 10] {};
        \node[main] (13) [right=2mm of 10] {};
        \node[main] (14) [below right=2mm of 10] {};

        \node[main] (20) [above right of=10] {};
        \node[main] (21) [below left=2mm of 20] {};
        \node[main] (23) [below right=2mm of 20] {};
        \node[main] (24) [below=2mm of 20] {};
        
        \node[main] (30) [below right of=20] {};
        \node[main] (31) [left=2mm of 30] {};
        \node[main] (32) [above left=2mm of 30] {};
        \node[main] (34) [below left=2mm of 30] {};
        
        \node[main] (40) [below right of=10] {};
        \node[main] (41) [above left=2mm of 40] {};
        \node[main] (42) [above=2mm of 40] {};
        \node[main] (43) [above right=2mm of 40] {};
        
        \draw (12) -- (21);
        \draw (13) -- (31);
        \draw (14) -- (41);
        \draw (23) -- (32);
        \draw (24) -- (42);
        \draw (34) -- (43);

        \begin{scope}[on background layer]     
            \node[party] () [right=-8mm of 10] {};
            \node[party] () [below=-8mm of 20] {};
            \node[party] () [left=-8mm of 30] {};
            \node[party] () [above=-8mm of 40] {};
        \end{scope}
        
    \end{tikzpicture}
    \caption{Privacy can be restored in a trivial way by allowing each pair of parties to establish pairwise correlations. This can be achieved if each party holds multiple nodes of a graph state, as illustrated here. If each party measures those of their nodes that are correlated with other nodes, they can use the outcomes to encode their remaining node in such a way that their secret parameter remains private.}
    \label{fig: restoring privacy}
\end{figure}

Similarly, the parties can restore privacy without a pre-shared secret if the resource state has more than mode per party. Specifically, consider the graph state represented in Fig.~\ref{fig: restoring privacy}. The parties could use the bipartite shared states to establish pairwise secrets (by measuring them), which could then be used to restore privacy. If we want to consider a ``one-step" process that simply involves a single encoding, we can construct a channel mapping from an $N$-mode state to a one-mode state that each party applies and that acts in the same way as measuring out the bipartite states and encoding the multipartite graph accordingly.

Nonetheless, this method of restoring privacy can be considered in some sense trivial, rather than a fundamental physical property of the resource, since we are layering privacy onto a non-private protocol via secret sharing. It is possible to learn $(\theta_i +\xi_i)$ perfectly; we have simply redefined the secret parameter as being connected to the physical parameter of the quantum state in a different way. Indeed, after the secret is shared between the parties, we no longer even need entanglement between the parties, since the privacy is purely classical. The QFI matrix also has more than a single non-zero eigenvalue if we take it with regard to the parameters $\{\xi_i\}$ as well as $\{\theta_i\}$. Further, one important reason we are interested in privacy in the first place is the link with sensitivity: in the DV case, GHZ-states both allow perfect privacy and an $\mathcal{O}[N]$ improvement in the sensitivity over all separable protocols (in terms of the QFI with regard to the function of interest). However, if we restore privacy using a classical secret, there is no such link. Hence, restoring privacy in this way is not so theoretically interesting. This motivates us to use a more ``physical" privacy measure, in terms of the reduction in the QFI with regard to one party's parameter when another party acts to hide it, which we call relative privacy.

\section{Relative privacy}

Although complete privacy is not possible with Gaussian resources, the precision with which we can estimate a global function can grow a lot quicker than the precision with which we can estimate any local parameter.
Hence, when trying to learn a global function of parameters to a certain precision, we may be able to give away a much smaller (but non-zero) amount of information about the individual parameter values. We can regard this as relative privacy, since one function of parameters is less well-known than another.

We can reformulate our notions of privacy from the start of Section~\ref{sec: conditions} in this new relative setting: instead of wanting there to be multiple sets of parameters that result in the same encoded state, we now want multiple sets of parameters to result in \textit{almost} the same encoded state. Each set of parameters that results in a given value of our target function should result in similar values of the ``hidden" functions, since the encoded state should be relatively uninformative about their values.
Similarly, instead of requiring that every pair of parties $(i,j)$ can choose their parameter values so that they ``cancel out" and act as the identity on the probe state, we now require that they \textit{almost} cancel out.

We will now formalise this latter condition. 
We want to quantify how well one party can hide the parameter value of another party. That is, given party $x$ has acted (locally) with an encoding unitary on the (here assumed to be pure) resource state, $| \psi \rangle$, and party $y$ has also acted locally on their mode of $| \psi \rangle$, how close to the identity can the overall unitary be? We define the ratio
\begin{equation}
    \mathcal{R}_{xy} = \min_{\hat{H}_x} \Bigg[ \min_{\hat{H}_y} \bigg[ \frac{\mathrm{QFI}_{\phi} \big[ e^{i\phi (\hat{H}_x \otimes \hat{H}_y \otimes \mathbf{I}_{\overline{xy}})} | \psi \rangle \big]}{\mathrm{QFI}_{\phi} \big[ e^{i\phi (\hat{H}_x \otimes \mathbf{I}_{\overline{x}})}  | \psi \rangle \big]} \bigg] \Bigg].
\end{equation}
The operational interpretation of this is that if all parties other than $x$ and $y$ are untrusted, we can still estimate some linear function of $x$ and $y$'s parameter values $\mathcal{R}_{xy}^{-1}$ times better than their individual parameter values. We then maximise $\mathcal{R}_{xy}$ over all parties $x$ and $y$, to find the worst case scenario.
This differs slightly from the privacy measure given in Ref.~\cite{bugalho_private_2025}. There privacy is defined for a particular encoding and choice of function, whereas we minimise over all encodings and functions. Ref~\cite{bugalho_private_2025} also uses the trace of the QFI matrix as the denominator, whereas our measure is in terms of how well a particular pair of parties can hide each other's values.

The following simplification exists:
\begin{equation}
    \mathcal{R}_{xy} = \min_{\hat{H}_x} \bigg[ \frac{\mathrm{QFI}_{\phi} \big[ e^{i\phi (\hat{H}_x \otimes \mathbf{I}_{\overline{x}})} \mathrm{Tr}_y [| \psi \rangle\langle \psi |] e^{-i\phi (\hat{H}_x \otimes \mathbf{I}_{\overline{x}})} \big]}{\mathrm{QFI}_{\phi} \big[ e^{i\phi (H_x \otimes \mathbf{I}_{\overline{x}})}  | \psi \rangle \big]} \bigg],
    \label{eq: ratio expression}
\end{equation}
where we have replaced the minimisation over $H_y$ with a trace over the $y$ mode. This follows from the fact that the QFI of a mixed state is the minimum over all purifications. Since we only need to carry out a single minimisation, this quantity should be easier to calculate.

We now note that a fixed (parameter-independent) unitary on the idler modes cannot change the QFI, so we are allowed to apply any unitary on the modes other than $x$ before calculating the QFI. Since any $m$-mode Gaussian state can be purified with no more than $m$ other modes, and all purifications are unitarily equivalent, we can reduce the calculation of $\mathrm{QFI}_{\phi} \big[ e^{i\phi (\hat{H}_x \otimes \mathbf{I}_{\overline{x}})}  | \psi \rangle \big]$ to calculating the QFI for a two-mode squeezed vacuum (TMSV), up to local unitaries on the $x$ mode. The numerator is slightly more complicated, since $\mathrm{Tr}_{\overline{xy}} [| \psi \rangle\langle \psi |]$ could, in theory, require two purifying modes, rather than one. It reduces to calculating the QFI of an (up to) three mode state (two idler modes) as opposed to a two mode state.

In other words, for any Gaussian initial resource, $| \psi \rangle$, the proportion of the parameter information encoded by party $x$ that \textit{cannot} be hidden by party $y$ is at least $\mathcal{R}_{xy}$. $\mathcal{R}_{xy}$ can be calculated as follows:
\begin{enumerate}
    \item Denote the average photon number of $\mathrm{Tr}_{\overline{x}} [| \psi \rangle\langle \psi |]$ (i.e., the average photon number of mode $x$) as $\bar{n}$.
    \item Denote the TMSV whose average photon number per mode is $\bar{n}$ as $|\Psi_{\bar{n}}\rangle$ (equivalently, this is the TMSV parametrised by squeezing parameter $r=\frac{1}{2}\mathrm{arccosh}[2\bar{n}+1]$). Label its two modes $S$ (signal) and $I$ (idler).
    \item For any $| \psi \rangle$ and pair of parties, $x$ and $y$, there exists a Gaussian noise channel, $\Phi_{| \psi \rangle , x,y}$, such that applying it to the idler mode of $|\Psi_{\bar{n}}\rangle$ results in a state that is identical to $\mathrm{Tr}_{y} [| \psi \rangle\langle \psi |]$ up to local unitaries (i.e., unitaries acting separately on mode $x$ and the set of modes $\overline{xy}$). $\Phi_{| \psi \rangle , x,y}$ could be either a single mode channel or a two-mode channel, in which case we define it as acting on both the idler mode and a vacuum mode. The subscripts communicate that the choice of channel depends on the particular $| \psi \rangle$ and pair of parties chosen.\footnote{Explicitly: let $| \psi \rangle$ consist of $N$ modes. We choose $\Phi_{| \psi \rangle , x,y}$ so that $\big(\mathbf{I}_{SB}\otimes\Phi_{IA}\big) \big[ |\Psi_{\bar{n}}\rangle_{SI} \otimes |0\rangle_{A} \otimes |0\rangle^{\otimes N-4}_{B} \big] \overset{u\otimes v}{=} \mathrm{Tr}_{y} [| \psi \rangle\langle \psi |]$, where $\overset{u\otimes v}{=}$ denotes equality up to some pair of unitaries acting separately on the $S$ mode and the $I$, $A$, and $B$ modes. The form of $\Phi$ depends on the initial state $| \psi \rangle$.}
    \item The ratio $\mathcal{R}_{xy}$ can now be calculated as
    \begin{equation}
        \mathcal{R}_{xy} = \min_{\hat{H}} \Bigg[ \frac{\mathrm{QFI}_{\phi} \big[ \big( e^{i\phi \hat{H}}\otimes \Phi_{| \psi \rangle , x,y} \big) \big[|\Psi_{\bar{n}}\rangle\big]  \big]}{\mathrm{QFI}_{\phi} \big[ \big( e^{i\phi \hat{H}}\otimes \mathbf{I} \big) \big[|\Psi_{\bar{n}}\rangle\big]  \big]} \Bigg],
    \end{equation}
    where the structure of the initial resource features only in the particular noise channel considered.
\end{enumerate}

We can therefore relate the proportion of information that cannot be hidden to the robustness of a TMSV to a particular noise channel. The remaining minimisation is still non-trivial, but can be bounded by restricting $\hat{H}$ to certain classes of encoding Hamiltonians. For instance, we can minimise over all Gaussian encoding unitaries.
Further, if we are able to lower bound the robustness of a TMSV to an idler noise channel, we will have a bound on the maximum achievable privacy for any Gaussian resource state, although we leave this for a future work.

\subsection{Example: Fully-connected graph state}

To illustrate what this process entails, we consider the example of a graph state resource on $N$ nodes formed by one-mode squeezing each node with strength $\nu$ and then applying a controlled-phase gate of strength $\mu$ between each pair of nodes. The covariance matrix of the resulting fully-connected graph state, $|\psi_{\mathrm{FC}}\rangle$, takes the block form\footnote{We use the convention that the quadratures are ordered $(\hat{q}_1,\hat{p}_1,\dots\hat{q}_N,\hat{p}_N)$, rather than $(\hat{q}_1,\dots\hat{q}_N,\hat{p}_1,\dots\hat{p}_N)$.}
\begin{align}
    &V_{\mathrm{FC}} =
    \begin{pmatrix}
        \mathbf{a} &\mathbf{b} &\dots &\mathbf{b}\\
        \mathbf{b} &\mathbf{a} & &\mathbf{b}\\
        \vdots & &\ddots &\vdots\\
        \mathbf{b} &\mathbf{b} &\dots &\mathbf{a}
    \end{pmatrix},\label{eq: Vfc}\\
    &\mathbf{a} =
    \begin{pmatrix}
        e^{2\nu} &0\\
        0 &e^{-2\nu}+(N-1)e^{2\nu}\mu^2
    \end{pmatrix},\\
    &\mathbf{b} =
    \begin{pmatrix}
        0 &e^{2\nu}\mu\\
        e^{2\nu}\mu &(N-2)e^{2\nu}\mu^2
    \end{pmatrix}.
\end{align}
For simplicity, we apply local squeezing and a phase rotation to each mode so that the covariance matrix is instead composed of diagonal blocks:
\begin{align}
    &V_{\mathrm{FC}}' =
    \begin{pmatrix}
        \bm{\alpha} &\bm{\beta} &\dots &\bm{\beta}\\
        \bm{\beta} &\bm{\alpha} & &\bm{\beta}\\
        \vdots & &\ddots &\vdots\\
        \bm{\beta} &\bm{\beta} &\dots &\bm{\alpha}
    \end{pmatrix},\\
    &\bm{\alpha} = \sqrt{1+\lambda^2(N-1)}\mathbf{I},\\
    &\bm{\beta} =
    \begin{pmatrix}
        \frac{(N-2)\lambda^2-\lambda\sqrt{4+N^2\lambda^2}}{2\sqrt{1+(N-1)\lambda^2}} &0\\
        0 &\frac{(N-2)\lambda^2+\lambda\sqrt{4+N^2\lambda^2}}{2\sqrt{1+(N-1)\lambda^2}}
    \end{pmatrix},
\end{align}
where we combine $\nu$ and $\mu$ into the parameter $\lambda = e^{2\nu}\mu$.

If an encoding unitary, $e^{i \theta\hat{H}}$, acts only on the first mode, the QFI is unaffected by a unitary acting on the other modes, so we are allowed to apply the symplectic transformation that diagonalises the $2(N-1)$ by $2(N-1)$ block forming the bottom-right of the covariance matrix. Hence, we can write the encoded graph state as
\begin{equation}
    |\psi_{\mathrm{FC},\mathrm{enc}}\rangle = \Big( e^{i \theta\hat{H}} u\otimes v \Big)\Big|\Psi_{\frac{1}{2}\big(\sqrt{1+\lambda^2(N-1)} - 1\big)}\Big\rangle ,
\end{equation}
where $u$ and $v$ are $\theta$-independent unitaries acting on the TMSV with covariance matrix
\begin{align}
    V_{\Psi} =
    \begin{pmatrix}
        \bm{\alpha} &\lambda\sqrt{N-1} \mathbf{Z}\\
        \lambda\sqrt{N-1} \mathbf{Z} &\bm{\alpha}
    \end{pmatrix}.
    \label{eq: V TMSV}
\end{align}
Applying a unitary to the first mode after it has already been encoded also cannot change the QFI, so we can replace $e^{i \theta\hat{H}} u$ with $u^\dagger e^{i \theta\hat{H}} u$. Then, the unitaries on the first mode simply act as a change of basis for the Hamiltonian, $\hat{H}$, and since, in Eq.~(\ref{eq: ratio expression}), we wish to minimise over all Hamiltonians, we can absorb the unitaries into the choice of $\hat{H}$.

If we trace out one mode from the fully-connected graph state, i.e., remove one row and column from the covariance matrix in Eq.~(\ref{eq: Vfc}), the covariance matrix of the resulting state can no longer be diagonalised via local unitaries to give Eq.~(\ref{eq: V TMSV}). Instead, diagonalising the bottom-right $2(N-2)$ by $2(N-2)$ block gives a state, $\Psi'$, with the covariance matrix\footnote{We follow the methodology of Ref.~\cite{serafini_unitarily_2005}, applying a unitary with matrix elements $u_{mn}=\frac{e^{imn\phi}}{\sqrt{N-2}}$, where $\phi=\frac{2\pi}{N-2}$.}
\begin{align}
    &V_{\Psi'} =
    \begin{pmatrix}
        \bm{\alpha} &\sqrt{N-2}\bm{\beta}\\
        \sqrt{N-2}\bm{\beta} &\bm{\alpha}+(N-3)\bm{\beta}
    \end{pmatrix}.
\end{align}
State $\Psi'$ is the TMSV $\Psi$ after the application of a one-mode Gaussian channel to a single mode. The one-mode Gaussian channels are characterised by their canonical forms~\cite{weedbrook_gaussian_2012}, and the particular channel that transforms $\Psi$ into $\Psi'$ can be identified, up to local Gaussian unitaries\footnote{Specifically, before applying the loss channel, we apply a one-mode squeezing operation to the second mode, with squeezing parameter $r=\log\Bigg[\frac{\sqrt{4+N^2\lambda^2}+(N-2)\lambda}{\sqrt{4+N^2\lambda^2}-(N-2)\lambda}\sqrt{\frac{2+N\lambda^2+\lambda\sqrt{4+N^2\lambda^2}}{2+N\lambda^2-\lambda\sqrt{4+N^2\lambda^2}}}\Bigg]$.}, with a pure loss channel whose transmissivity is $\frac{N-2}{N-1}$.

Since every pair of nodes is equivalent to every other pair, we can calculate the maximum degree to which one party can hide another party's individual parameter value as the ratio
\begin{equation}
    \mathcal{R}_{xy} = \min_{\hat{H}} \left[ \frac{\mathrm{QFI}_{\phi} \big[ \big( e^{i\phi \hat{H}}\otimes \mathbf{I} \big) \big[ \Psi_{\frac{1}{2}\big(\sqrt{1+\lambda^2(N-1)} - 1\big)} \big]  \big]}{\mathrm{QFI}_{\phi} \big[ \big( e^{i\phi \hat{H}}\otimes \mathcal{L} \big) \big[ \Psi_{\frac{1}{2}\big(\sqrt{1+\lambda^2(N-1)} - 1\big)} \big]  \big]} \right],
\end{equation}
where $\mathcal{L}$ denotes the pure loss channel along with the local Gaussian unitaries. Hence, the proportion of information encoded by a particular Hamiltonian that can be hidden by another party is equivalent to the reduction in sensitivity of a TMSV probe of that same Hamiltonian after the application of a loss channel to the idler mode.

\section{Discussion}

We have shown, via a simple no-go theorem, that perfect privacy cannot be achieved by any distributed sensing protocol that uses Gaussian states as a resource. This holds regardless of whether the encoding channels are unitary, non-unitary, Gaussian, or non-Gaussian.

Privacy can be restored in a trivial way using a pre-shared classical secret, however this does not result in enhanced sensitivity and is theoretically uninteresting. Therefore, we introduce a measure of relative privacy, in terms of how well any party can hide another party's parameter value, and discuss how it could be calculated.
An interesting line for future research would be to bound this privacy measure for all Gaussian states, and so to find the ultimate limit on achievable privacy using Gaussian resources.

\bigskip
\noindent\textbf{Acknowledgments}\quad~The authors acknowledge financial support from the ANR, through the projects PEPR-HQI (award number ANR-22-PNCQ0002) and EQUINE (ANR-23-QUAC-0001), and from the European Union's Horizon Europe programme, through the project CLUSTEC (grant agreement No.~101080173).

\bibliography{bibliography}

\end{document}